\documentstyle[12pt]{ioplppt}
\input tcilatex

\QQQ{Language}{
American English
}

\begin{document}

\title{The Distribution of Multiple Energy Minima and Hysteresis in Zero
Temperature Random Field Ising Model}
\author{A.A.Likhachev}

\begin{abstract}
The Random Field Ising Model (RFIM) is the simplest physical model
reflecting effect of quenched disorder on the different types of phase
transitions in solids. The presence of multiple energy minima in the RFIM is
an important feature determining main physical properties of the systems
with a field type of quenched disorder. In particular, according to computer
simulations of Sethna at al. performed for a zero temperature RFIM, the
irreversible loop of magnetization appears as direct result of their
presence. In the present paper the analytic method based on the total number
of energy minima estimation is developed to study the hysteretic behavior of
a zero temperature RFIM. As a result, the magnetization hysteresis loop is
obtained as a marginal curve separating metastable and completely unstable
states in the magnetization-external field plane.
\end{abstract}

\address{Institute of Metal Physics, 36 Vernadsky St., 252142 Kiev, Ukraine}


\section{Introduction}

The Random Field Ising Model (RFIM) is the simplest physical model
reflecting the influence of different type of defects or quenched disorder
on the different types of phase transitions in solids. Starting from the
pioneering work of Imry and Ma \cite{a1} a great number of investigations
have been made to understand the nature of different phase transitions in
presence of random bond or field type of quenched disorder \cite
{a2,a3,a4,a5,a6,a7,a8}. The literature contains several reviews on the
theory and experiments \cite{a9,a10,a11,a12,a13,a14,a15}. The investigations
developed in time from the equilibrium properties and lower critical
dimensionality to dynamics of domain walls, metastability and hysteretic
properties. The RFIM is now also used to explain a role of impurities in
displacive transitions and random fields effect in charge-density waves. The
multiple energy minima in the RFIM is an important feature determining main
equilibrium and non-equilibrium physical properties of these systems. The
most important effect resulting from presence of the infinite number of
local energy minima in RFIM can be expected out of equlibrium (for example,
in zero temperature case) when the thermal fluctuations become negligibly
small.In particular, the irreversible magnetization loop appears as direct
result of multiple energy minima in RFIM. This is confirmed by the direct
computer simulations of magnetization processes reported in several
publications (see \cite{a16,a17}). Due to great number of possible
stationary microscopic configuration the magnetization process can be
considered as an infinite bifurcation sequence of transitions between the
multiple stationary (metastable) states when the external field is changed.
Such a transitions occur each time as the external field approaches the
corresponding critical value where the current spin configuration becomes
unstable. Then, this unstable configuration is changed at a fixed external
field through a sequence of spin-by-spin flip transitions until a new
stationary state to be achieved. The main topic of this paper is to
represent the alternative analytic method that gives a possibility to
establish the direct connection between the infinite number of metastable
states in RFIM and magnetization hysteresis loop. The method is based on the
calculations of the total number of energy minima for the random field Ising
Hamiltonian corresponding to different fixed values of the external field
and specific magnetization per site. The hysteresis looop appears as a
marginal curve separating $m-h$ plane into the metastability region inside
of the loop and the region of completely unstable states outside of it.

\section{Local Energy Minima in RFIM}

\subsection{Stationary States of Zero Temperature RFIM Hamiltonian}

Consider the Ising spin system $\sigma _i=\pm 1$ defined on a regular $N$%
-site lattice and interacting with the nearest neighbors, external uniform
fild $h$ and irregular $\tau _i$ (random) fields (here, $i\in 1,...,N$ is a
lattice site number). The energy of such a system is known to be represented
as follows: 
\begin{equation}
\label{eq1}H\{{\bf s}\}=-\frac 12J\stackunder{ij}{\sum }v_{ij}\sigma
_i\sigma _j-\stackunder{i}{\sum }\left( h+\tau _i\right) \sigma _i 
\end{equation}
where, the first term represents the nearest neighbors interaction energy ($%
v_{ij}=1\neq 0$ only for nearest neighbors), $J$ is a spin-spin interaction
constant and the second one reflects the influence of both the external and
the random fields; $\left\{ {\bf s}\right\} {\bf \equiv }\left\{ \sigma
_1,\sigma _2,...,\sigma _N\right\} $ denotes $N$-spin configuration vector.
The random field values $\tau _i$ are assumed to be non-correlated at the
different lattice sites and characterized with the probability distribution
function $P\left( \tau \right) $ that is the same for all the lattice sites.
One can also define the possible local minima of RFIM Hamiltonian as a set
of Ising spin configurations ${\bf s}$ satisfying the following conditions: 
\begin{equation}
\label{eq2}H\{{\bf s}\}<H\{{\bf s}^{\left( i\right) }\}\,\,\,\,\,\,\forall
i\in 1,...,N 
\end{equation}
where, each local minimum ${\bf s}$ is a spin configuration having lower
energy in respect to all other $N$ configurations ${\bf s}^{\left( i\right)
} $ that differ from the initial one ${\bf s}$ by a single spin at $i$-th
site flipped in opposite direction: $\sigma _i\rightarrow -\sigma _i$. In
other words, the system of inequalities: 
\begin{equation}
\label{eq3}\fl u_i\{{\bf s}\}=\frac 12[H\{{\bf s}^{\left( i\right) }\}-H\{%
{\bf s}\}]=\sigma _i\left( \tau _i+h+J\stackunder{j}{\sum }v_{ij}\sigma
_j\right) >0;\,\,\,\,\,\,\,\,\forall i\in 1,...,N 
\end{equation}
define all the possible stationary spin configurations for zero temperature
RFIM. It should be noted that these inequalities can be satisfied in the
unique case when both factors in the product $\sigma _i\left( \tau _i+h+J%
\stackunder{j}{\sum }v_{ij}\sigma _j\right) $will have the same signs.
Therefore, the solution of inequalities is completely equivalent to the
following equations: 
\begin{equation}
\label{eq3a}\sigma _i=sign\left( \tau _i+h+J\stackunder{j}{\sum }%
v_{ij}\sigma _j\right) 
\end{equation}
One can interpret these equations as local equilibrium conditions for each
site spin in the local self-consistent field 
\begin{equation}
\label{eq3b}h_i=\tau _i+h+J\stackunder{j}{\sum }v_{ij}\sigma _j 
\end{equation}
Accordingly, each site spin is oriented such a way to have the same
direction as the local field applied. This equation also represents a zero
temperature limit of the mean field equation that is often used in different
studies of the equilibrium properties of RFIM. In particular case $J=0$ the
Eq.(\ref{eq3a}) immediately gives the only possible solution: 
\begin{equation}
\label{eq3c}\sigma _i=sign\left( \tau _i+h\right) 
\end{equation}
that corresponds to a single minimum of RFIM in this case. The total
magnetization per site in this case can be expressed as 
\begin{equation}
\label{eq3d}m_0(h)=\frac 1N\stackunder{j}{\sum }\sigma _i=\int d\tau P(\tau
)sign(h+\tau ) 
\end{equation}
where, denoting the Dirac's function as $\delta (t)$ one can define the
probability distribution function of the random field values: 
\begin{equation}
\label{eq3e}P(t)=\frac 1N\stackunder{i}{\sum }\delta (t-\sigma _i) 
\end{equation}
$m_0(h)$ will therefore represent a single magnetization curve in case of
non-interacting Ising spin system in a random field.

\subsection{Numerical Simulation of Magnetization Curves}

The most important effect resulting from presence of the infinite number of
local energy minima in RFIM can be expected for the magnetization properties
of Ising spin system in zero temperature case when thermal fluctuations do
not play any role. In absence of thermal fluctuations each stationary state
will still remain the same as long as the first external field value is
achieved where this state becomes unstable and pinning equations Eq.(\ref
{eq3}) or Eq.(\ref{eq3a}) can not be more satisfied. Then, the unstable spin
configuration is changed at fixed external field through a sequence of
single spin flip transitions until a new stationary state to be achieved.
Therefore, due to great number of possible stationary microscopic
configuration consistent with the pinning conditions the magnetization
process can be considered as a bifurcation sequence of transitions between
the different stationary states as the external field is changed. The
necessary condition for a single spin transition can be represented as
follows: 
\begin{equation}
\label{eq3f}\sigma _i\Longrightarrow -\sigma
_i\,\,\,\,\,\,if\,\,\,only\,\,\,\,u_i\{{\bf s}\}<0\,\,\,\,\,\&\,\,\,\,\,u_i\{%
{\bf s}\}=\min 
\end{equation}
It is denotes that new stable spin configuration at each fixed external
field is achieved through a sequence spin by spin flip processes running
along a single possible configuration path providing the maximal energy gain
at each elementary stage of transition. The configuration path satisfying
conditions Eq.(\ref{eq3f}) provides the maximal rate of the free energy
decrease in agreement with usual requirements of non-linear thermodynamics
and give necessary algorithm for the computer simulation methods and
modeling of magnetization processes in framework of RFIM. Fig.1 shows the
typical magnetization loop resulting from such a type simulation procedure
performed for the one-dimensional RFIM. Numerically simulated gaussian
random field values on the one-dimensional lattice with 1000 sites were used
for calculations.

\section{Analytic Accounting of the number of local minima}

\subsection{General expressions}

By using the definition from Eq.(\ref{eq3})one can propose the following
general way for evaluation of the total number of local energy minima in
RFIM. For this aim, denote initially the half of energy difference as: And
then, one can easily construct the characteristic functional containing
information on the all possible local minima in RFIM that can be written as
follows: 
\begin{equation}
\label{eq4}L\{{\bf s}\}=\stackunder{i}{\prod }\eta \left( u_i\{{\bf s}%
\}\right) 
\end{equation}
where, $\eta \left( t\right) $ is the Heaviside function ($\eta \left(
t\right) =1$ , if $t>0$ and $\eta \left( t\right) =0$, if $t<0$). One can
easily check that the above introduced characteristic functional takes the
unit value only if any spin configuration is a local minimum of RFIM
Hamiltonian and zero in other case. Therefore, taking the sum over all
possible $2^N$ Ising spin configurations one can obtain the following
general representation for the total number of local energy minima in RFIM: 
\begin{equation}
\label{eq5}Z\left( h\right) =\stackunder{\{{\bf s}\}}{\sum }L\{{\bf s}\} 
\end{equation}
Respectively, one can also find the averaged over the random field ensemble
number of local minima as follows: 
\begin{equation}
\label{eq6}\overline{Z\left( h\right) }=\stackunder{\{{\bf s}\}}{\sum }%
\left\langle L\{{\bf s}\}\right\rangle _{{\bf \tau }} 
\end{equation}
Due to statistical independence of the random field values $\tau _i$ for
different lattice sites the averaging procedure in Eq.(\ref{eq6}) is
transformed into the product of one-site averages: 
\begin{equation}
\label{eq7}\overline{Z\left( h\right) }=\stackunder{\{{\bf s}\}}{\sum }%
\stackunder{i}{\prod }\left( \frac 12+\frac 12\sigma _im_0\left( h+J%
\stackunder{j}{\sum }v_{ij}\sigma _j\right) \right) 
\end{equation}
Here, we used the following evident property of Heaviside function: $\eta
\left( t\right) +\eta \left( -t\right) =1$ and $\eta \left( t\right) -\eta
\left( -t\right) =sign\left( t\right) $ and also $\eta \left( \sigma
_it\right) =\frac 12\left( 1+\sigma _isign\left( t\right) \right) $ where, 
\begin{equation}
\label{eq8}m_0\left( t\right) =\int sign\left( t+\tau \right) P\left( \tau
\right) d\tau . 
\end{equation}
It will be shown below that $m_0\left( t\right) $ represents a magnetization
curve of RFIM in case of non-interacting spins when $J=0$. Therefore, the
problem of the total local minima number estimation for the Random field
Ising model can be always reduced to a standard statistical-mechanical
procedure on calculation of the partition function for a generalized Ising
model: 
\begin{equation}
\label{eq9}\overline{Z\left( h\right) }=\stackunder{\{{\bf s}\}}{\sum }\exp
\left( V\{{\bf s}\}\right) 
\end{equation}
with some effective Hamiltonian: 
\begin{equation}
\label{eq10}V\{{\bf s}\}=-N\ln 2+\stackunder{i}{\sum }\ln \left( 1+\sigma
_im_0\left( h+J\stackunder{j}{\sum }v_{ij}\sigma _j\right) \right) 
\end{equation}
It is important that the effective Hamiltonian can be always expressed as a
finite order polynomial of Ising spin variables: 
\begin{equation}
\label{eq11}V\{{\bf s}\}=V^0+\stackunder{i}{\sum }V_i^1\sigma _i+\stackunder{%
ij}{\sum }V_{ij}^2\sigma _i\sigma _j+\stackunder{ijk}{\sum }V_{ijk}^3\sigma
_i\sigma _j\sigma _k+... 
\end{equation}
Polynomial coefficients are translation invariant functions of their
arguments. These take zero values for coinciding pairs of their arguments
and contain only the short range nearest neighbor interactions. The maximal
order of this polynomial is $k+1$ where, $k$ is a number of nearest
interacting neighbors of the initial RFIM Hamiltonian. As a result, using
the analogy with the partition function of equilibrium thermodynamics one
can assume the following universal asymptotic behavior for the average
number of local minima when the lattice sites number $N$ becomes infinitely
large: 
\begin{equation}
\label{eq12}\overline{Z\left( h\right) }\stackunder{N\rightarrow \infty }{%
\rightarrow }\exp \left( \mu \left( h\right) N\right) 
\end{equation}
where $\mu \left( h\right) $ is the independent on $N$ function of the
external field. Because the total number of local energy minima $1\leq
Z\left( h\right) \leq 2^N,$ then $0\leq \mu \left( h\right) \leq \ln 2$ and
one can make a general conclusion that $\overline{Z\left( h\right) }$ must
exponentially grow as $N\rightarrow \infty $. Therefore, RFIM can have a
single minimum if only $\mu \left( h\right) \equiv 0$ and infinite number of
local minima in other case. Such a case is realized for the non-interacting
spin system at $J=0$. One can easily obtain this result by using Eq.(\ref
{eq7}) at $J=0$: 
\begin{equation}
\label{eq13}\overline{Z\left( h\right) }_{J=0}=\left( \frac 12\stackunder{%
\sigma =\pm 1}{\sum }\left( 1+\sigma m_0\left( h\right) \right) \right) ^N=1 
\end{equation}
Naturally, in general case both $\overline{Z\left( h\right) }$ and $\mu
\left( h\right) $ are the functions dependent on the parameters of Random
Field Ising Hamiltonian such as the external field $h$, spin-spin
interaction constant $J$ and also on the distribution function of random
field $P(\tau )$. Along with the total number of local energy minima $%
\overline{Z\left( h\right) }$ one can also consider another important
quantity: $\overline{Z\left( M,h\right) }$ that gives the number of
stationary states corresponding to a fixed value of total magnetization $M=%
\stackunder{i}{\sum }\sigma _i$. Similar to Eq.(\ref{eq9}) this one can be
found as follows: 
\begin{equation}
\label{eq14}\overline{Z\left( M,h\right) }=\stackunder{\{{\bf s;}M\}}{\sum }%
\exp \left( V\{{\bf s}\}\right) 
\end{equation}
where, the summation is performed over configurations with a fixed total
magnetization value. One can also expect the following universal asymptotics
in the thermodynamical limit $N\rightarrow \infty $: 
\begin{equation}
\label{eq15}\overline{Z\left( M,h\right) }\longrightarrow \exp \left( \mu
\left( m,h\right) N\right) 
\end{equation}
where, $m=M/N$ is a specific magnetization per site. But unlike to always
positive function $\mu (h)$ defined above the new function $\mu \left(
m,h\right) $ introduced here and dependent on two variables $m$ and $h$ can
take the negative values too in the corresponding area of their change. As a
result of Eq.(\ref{eq15}), the number of possible stationary states
corresponding to different fixed values of specific magnetization $m$
becomes infinitely small approaching zero value as $N\rightarrow \infty $\
in that region of $m-h$ plane where $\mu \left( m,h\right) <0$. Therefore,
one can conclude that all possible infinite set of stationary solutions is
always located only inside the region $\mu \left( m,h\right) >0$. The
marginal curve $\mu \left( m,h\right) =0$ separating these two regions can
be therefore associated with a global magnetization loop. Such
interpretation is completely consistent with the results of dynamical
approach based on the computer simulation of partial magnetization sub-loops
that are always locked inside of main loop confirming its marginal character
as shown in Fig.1.

Due to above mentioned analogy with the equilibrium statistical mechanics a
great number of approximation methods can be applied to calculation of $%
\overline{Z\left( h\right) }$ and $\overline{Z\left( M,h\right) }$. As an
example, the simple perturbation method will be used below to estimate the
number of local minima in case of the small interaction constant limit . The
main reason of that approach is to show that the infinite number of local
minima effect immediately appears as only $J\neq 0$, starting from an
arbitrary weak spin-spin interaction.

\subsection{Perturbation Theory Expansion}

As known the perturbation theory methods in statistical mechanics providing
correct behavior in the thermodynamic limit when $N\rightarrow \infty $ are
realized as a free energy expansion that is proportional to the log value of
a partition functional. Similarly, the corresponding expansion in respect to 
$\ln \left( \overline{Z\left( h\right) }\right) =N\mu \left( h\right) $ and $%
\ln \left( \overline{Z\left( M,h\right) }\right) =N\mu \left( M,h\right) $
should be performed in case $J\rightarrow 0$. Accordingly, one can easily
obtain the Taylor's series expansion for the effective Hamiltonian defined
in Eq.(\ref{eq10}) as $V\{{\bf s}\}=V_0\{{\bf s}\}+V_1\{{\bf s}\}+V_2\{{\bf s%
}\}$ up to the second order on $J$. Then, using Ising spin algebra rule ($%
\sigma ^2=\sigma $), a zero order term can be expressed as follows: 
\begin{equation}
\label{eq16}\fl V_0\{{\bf s}\}=\stackunder{i}{\sum }\ln \left( \frac{%
1+\sigma _im_0\left( h\right) }2\right) =\frac 12\ln \left( \frac{%
1-m_0^2\left( h\right) }4\right) N+\frac 12\ln \left( \frac{1+m_0\left(
h\right) }{1-m_0\left( h\right) }\right) \stackunder{i}{\sum }\sigma _i 
\end{equation}
Others first and second order perturbation contributions are: 
\begin{equation}
\label{eq17}\fl V_1\{{\bf s}\}=J\stackunder{i}{\sum }\frac{m_0^{^{\prime
}}\left( h\right) \sigma _i}{1+\sigma _im_0\left( h\right) }\stackunder{j}{%
\sum }v_{ij}\sigma _j=J\frac{m_0^{^{\prime }}\left( h\right) }{1-m_0^2\left(
h\right) }\stackunder{ij}{\sum }v_{ij}\left( \sigma _i-m_0\left( h\right)
\right) \sigma _j 
\end{equation}
\begin{eqnarray}
\label{eq18}
\fl V_2\{{\bf s}\}=-\frac 12J^2\stackunder{i}{\sum }\frac{\left(
m_0^{^{\prime }}\left( h\right) \right) ^2\sigma _i\sigma _i}{\left(
1+\sigma _im_0\left( h\right) \right) ^2}\left( \stackunder{j}{\sum }%
v_{ij}\sigma _j\right) ^2+\frac 12J^2\stackunder{i}{\sum }\frac{m_0^{^{\prime \prime }}\left( h\right) \sigma _i}{1+\sigma _im_0\left(
h\right) }\left( \stackunder{j}{\sum }v_{ij}\sigma _j\right) ^2=\nonumber \\ -\frac
12J^2
\frac{\left( m_0^{^{\prime }}\left( h\right) \right) ^2}{\left(
1-m_0^2\left( h\right) \right) ^2}\stackunder{ijk}{\sum }v_{ij}v_{ik}\left(
\sigma _i-m_0\left( h\right) \right) ^2\sigma _j\sigma _k+\nonumber \\ \frac 12J^2%
\frac{m_0^{^{\prime \prime }}\left( h\right) }{1-m_0^2\left( h\right) }%
\stackunder{ijk}{\sum }v_{ij}v_{ik}\left( \sigma _i-m_0\left( h\right)
\right) \sigma _j\sigma _k
\end{eqnarray}
As a result, the corresponding expansion of both the $\ln \left( \overline{%
Z\left( h\right) }\right) =N\mu \left( h\right) $ and $\ln \left( \overline{%
Z\left( M,h\right) }\right) =N\mu \left( m,h\right) $ up to the second order
on the interaction constant $J$ will look like as follows: 
\begin{equation}
\label{eq19}\ln \left( \overline{Z\left( h\right) }\right) =\ln \left( 
\overline{Z_0\left( h\right) }\right) +\left\langle V_1\right\rangle
_0+\left\langle V_2\right\rangle _0+\frac 12\left( \left\langle
V_1^2\right\rangle _0-\left\langle V_1\right\rangle _0^2\right) 
\end{equation}
\begin{equation}
\label{eq20}\ln \left( \overline{Z\left( M,h\right) }\right) =\ln \left( 
\overline{Z_0\left( M,h\right) }\right) +\left\langle V_1\right\rangle
_M+\left\langle V_2\right\rangle _M+\frac 12\left( \left\langle
V_1^2\right\rangle _M-\left\langle V_1\right\rangle _M^2\right) 
\end{equation}
Here, the averaging procedures are defined as: 
\begin{equation}
\label{eq21}\left\langle ...\right\rangle _0=\frac{\stackunder{\left\{ {\bf s%
}\right\} }{\sum }\left( ...\right) \exp V_0\{{\bf s}\}}{\stackunder{\left\{ 
{\bf s}\right\} }{\sum }\exp V_0\{{\bf s}\}}\,;\,\,\,\,\,\,\,\,\,\,\,\,\,\,%
\,\,\,\,\,\,\,\,\,\,\,\,\,\,\,\,\,\,\,\left\langle ...\right\rangle _M=\frac{%
\stackunder{\left\{ {\bf s,}M\right\} }{\sum }\left( ...\right) \exp V_0\{%
{\bf s}\}}{\stackunder{\left\{ {\bf s,}M\right\} }{\sum }\exp V_0\{{\bf s}\}}
\end{equation}
where, $\{{\bf s,}M\}$ is a set of spin configurations with a fixed
magnetization value $M$. Due to statistical independence of spin variables
in the different sites and evident result: $\left\langle \sigma
_i\right\rangle _0=m_0(h)$ that can be easily found from the Eqns.(\ref{eq16}%
, \ref{eq21}) both the $\left\langle V_1\right\rangle _0$ and second term
from the Eq.(\ref{eq18}) contributing into $\left\langle V_2\right\rangle _0$
take zero values.Taking also into account that acorrding to Eq.(\ref{eq13}) $%
Z_0(h)=1$ and $\ln \left( Z_0(h)\right) =0$ one can obtain the following
results for the total number of stationary states $\overline{Z\left(
h\right) }$:%
\begin{eqnarray}
\label{eq22}\fl\ln \left(\overline{ Z(h)}\right) =-\frac 12J^2\frac{\left(
m_0^{^{\prime }}\left( h\right) \right) ^2}{\left( 1-m_0^2\left( h\right)
\right) ^2}\stackunder{ijk}{\sum }v_{ij}v_{ik}\left\langle \left( \sigma
_i-m_0\left( h\right) \right) ^2\sigma _j\sigma _k\right\rangle _0+\nonumber\\\frac
12J^2\frac{\left( m_0^{^{\prime }}\left( h\right) \right) ^2}{\left(
1-m_0^2\left( h\right) \right) ^2}\stackunder{ijlk}{\sum }%
v_{ij}v_{lk}\left\langle \left( \sigma _i-m_0\left( h\right) \right) \left(
\sigma _l-m_0\left( h\right) \right) \sigma _j\sigma _k\right\rangle _0
\end{eqnarray}
The first term in square braces is completely reduced with a part $l=i$ of
the second sum. One can also check that only the $k=i\,\&\,l=j$ terms from
the remaining part of the second sum will give nontrivial contributions.
Therefore, 
\begin{eqnarray}
\fl\label{eq23}\ln \left(\overline{ Z(h)}\right) =\frac 12J^2\frac{\left( m_0^{^{\prime
}}\left( h\right) \right) ^2}{\left( 1-m_0^2\left( h\right) \right) ^2}%
\stackunder{ij}{\sum }\left( v_{ij}\right) ^2\left\langle 1-\sigma
_im_0\left( h\right) \right\rangle _0\left\langle 1-\sigma _jm_0\left(
h\right) \right\rangle _0=\nonumber\\\frac 12J^2\left( m_0^{^{\prime }}\left( h\right)
\right) ^2\stackunder{ij}{\sum }\left( v_{ij}\right) ^2=\frac 12J^2\left(
m_0^{^{\prime }}\left( h\right) \right) ^2kN
\end{eqnarray}
where $k$ is the number of nearest neighbors. Therefore, the total number of
energy minima becomes infinitely large even in case of an arbitrary weak
Ising spin interaction and can be expressed in the second order perturbation
theory approach as follows: 
\begin{equation}
\label{eq24}\overline{Z\left( h\right) }=\exp \left( \mu \left( h\right)
N\right) \,\,\,\,\,\,where,\,\,\,\,\,\,\,\,\mu \left( h\right) =\frac
12kJ^2\left( m_0^{^{\prime }}\left( h\right) \right) ^2 
\end{equation}

\subsection{Multiple minima and magnetization loop}

In this section the relationship between the great number of possible
stationary state of RFIM and the irreversible magnetization loop typical for
ferromagnetic materials will be a subject for study and discussion. It will
be shown that very important information on the magnetization loop can be
obtained not only from the direct dynamic simulations of the magnetization
process but also by accounting the number of possible stationary solutions
and their distribution on the different values of macroscopic magnetization.
More exactly, the method is based on the calculation of $Z\left( M,h\right) $
and solution of characteristic equation for the hysteresis loop: $\ln \left( 
\overline{Z(M,h)}\right) =\mu \left( m,h\right) N=0$. In the framework of
perturbation theory $\mu \left( m,h\right) $ can be found directly from the
Eq.(\ref{eq20}). But there is a simplest way to do that by using the evident
sum rule: $\overline{Z\left( h\right) }=\stackunder{M}{\sum }\overline{%
Z\left( M,h\right) }$ following from definitions of both these quantities.
Using this rule and exponential representations (see Eqns.(\ref{eq12}) and (%
\ref{eq15})) one can obtain asymptotically (as $N\rightarrow \infty $ ) the
following results: 
\begin{equation}
\label{eq25}\exp \left( \mu \left( h\right) N\right) =\stackunder{M}{\sum }%
\exp \left( \mu \left( m,h\right) N\right) \longrightarrow \exp \left( 
\stackunder{m\in [-1,1]}{\max }\left( \mu \left( m,h\right) \right) N\right) 
\end{equation}
Or, equivalent relationship: 
\begin{equation}
\label{eq26}\mu \left( h\right) =\mu \left( \widetilde{m}\left( h\right)
,h\right) \,\,\,\,\,\,\,\,\,where,\,\,\,\,\,\,\,\,\frac \partial {\partial
m}\mu \left( m,h\right) =0\,\,\,\,\,\,\,\,\,at\,\,\,\,\,\,m=\widetilde{m}%
\left( h\right) 
\end{equation}
As a result $\mu \left( m,h\right) $ can be expanded near the maximum $\,m=%
\widetilde{m}\left( h\right) $ as follows: 
\begin{equation}
\label{eq27}\mu \left( m,h\right) =-\frac 12\beta \left( h\right) \left( m-%
\widetilde{m}\left( h\right) \right) ^2+\mu \left( h\right)
;\,\,\,\,\,\,\,\,\,\,\,\,\,\,\,where,\,\,\,\,\,\,\,\,\,\beta \left( h\right)
>0
\end{equation}
In the case of weak interaction $J\rightarrow 0$ this representation can be
used to find the magnetization curve as a solution of the equation $\mu
\left( m,h\right) =0$. According to Eq.(\ref{eq24}) $\mu \left( h\right) =0$
at $J=0$ and there is a single degenerated solution $m=\widetilde{m}_0\left(
h\right) $for a zero order equation: 
\begin{equation}
\label{eq28}\mu _0\left( m,h\right) =-\frac 12\beta _0\left( h\right) \left(
m-\widetilde{m}_0\left( h\right) \right) ^2=0
\end{equation}
So, there is no hysteresis at $J=0$. But, as follows from the Eq.(\ref{eq27}%
), it is immediately appears as only $J\neq 0$ and represented by two
branches $m=m_{\pm }\left( h\right) $ of hysteresis loop close to $m=%
\widetilde{m}\left( h\right) $ and each other, where: 
\begin{equation}
\label{eq29}m_{\pm }\left( h\right) =\widetilde{m}\left( h\right) \pm \sqrt{%
2\mu \left( h\right) /\beta \left( h\right) };\,\,\,\,\,\,\,\,\,\,\,\,\,\,%
\,where,\,\,\,\,\,\,\,\,\,\beta \left( h\right) >0
\end{equation}
Because $\sqrt{\mu \left( h\right) }$ is the first order on $J$ then only up
to the first order calculations for $\widetilde{m}\left( h\right) =%
\widetilde{m}_0\left( h\right) \widetilde{+m}_1\left( h\right) $ and zero
order for $\beta \left( h\right) =\beta _0\left( h\right) $ should be
performed. In particular, one can easily show that $\widetilde{m}\left(
h\right) $ is exactly equal to a single-site average of Ising spin value $%
\left\langle \sigma _l\right\rangle $ and expanded into the perturbation
series as: 
\begin{equation}
\label{eq30}\widetilde{m}\left( h\right) =\left\langle \sigma
_l\right\rangle =\left\langle \sigma _l\right\rangle _0+\left\langle (\sigma
_l-\left\langle \sigma _l\right\rangle _0)V_1\left( {\bf s}\right)
\right\rangle _0=m_0\left( h\right) +kJm_0^{^{\prime }}\left( h\right)
m_0\left( h\right) 
\end{equation}
$\beta _0\left( h\right) $ can be found from the direct calculation of zero
order term $\mu _0\left( m,h\right) =$ $\frac 1N\ln \left( \overline{Z_0(M,h)%
}\right) $according to Eq.(\ref{eq14}) at $J=0$ and its expansion near the
maximum at $m=m_0\left( h\right) $.Therefore, $\beta _0\left( h\right)
=\left( -\partial ^2/\partial m^2\right) \mu _0\left( m,h\right) _{\left|
m=m_0\left( h\right) \right) }.$ Results of such calculations are
represented below: 
\begin{equation}
\label{eq31}\overline{Z_0\left( M,h\right) }=\stackunder{\{{\bf s;}M\}}{\sum 
}\exp \left( V_0\{{\bf s}\}\right) 
\end{equation}
\begin{eqnarray}
\fl\label{eq32}\mu _0\left( m,h\right) =\frac 1N\ln \left( \overline{Z_0(M,h)}%
\right) =\nonumber\\\frac 12[\left( 1+m\right) \ln \left( \frac{1+m_0\left( h\right) }{1+m}\right) +\left( 1-m\right) \ln \left( \frac{1-m_0\left( h\right) }{1-m}%
\right) ] 
\end{eqnarray}
\begin{equation}
\label{eq33}\beta _0\left( h\right) =\left( -\partial ^2/\partial m^2\right)
\mu _0\left( m,h\right) _{\left| m=m_0\left( h\right) \right) }=\frac
1{1-m_0^2\left( h\right) }
\end{equation}
The final equation representing the irreversible magnetization loop for a
zero temperature RFIM in the weak interaction limit will look like as
follows: 
\begin{equation}
\label{eq34}m_{\pm }\left( h\right) =\left( 1+kJm_0^{^{\prime }}\left(
h\right) \right) m_0\left( h\right) \pm Jm_0^{^{\prime }}\left( h\right) 
\sqrt{k\left( 1-m_0^2\left( h\right) \right) }
\end{equation}
A set of hysteresis loops calculated according to Eq.(\ref{eq34}) for
different values of the interaction parameter $J$ is indicated in Fig.2.

Therefore, the present approach makes it possible to perform detalied
anlytic studies concerning physical nature of hysteresis by using the Random
Field Ising based microscopic consideration.

References

Figure cuptions

Fig.1 Magnetization loop and partial sub-loops numerically simulated for the
one-dimentional RFIM.

Fig.2 Marginal hysteresis loops separating metastable band unstable states
in RFIM calculated for the different values of interaction constant in the
framework of perturbation method.

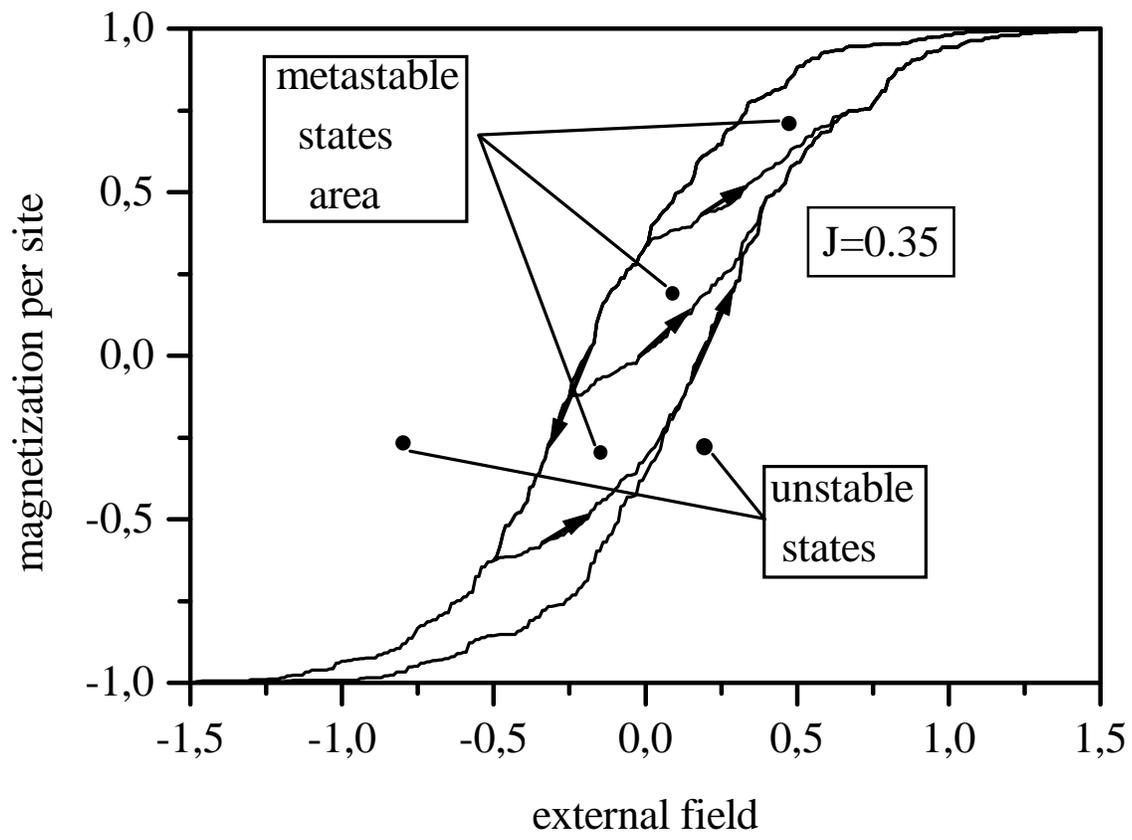
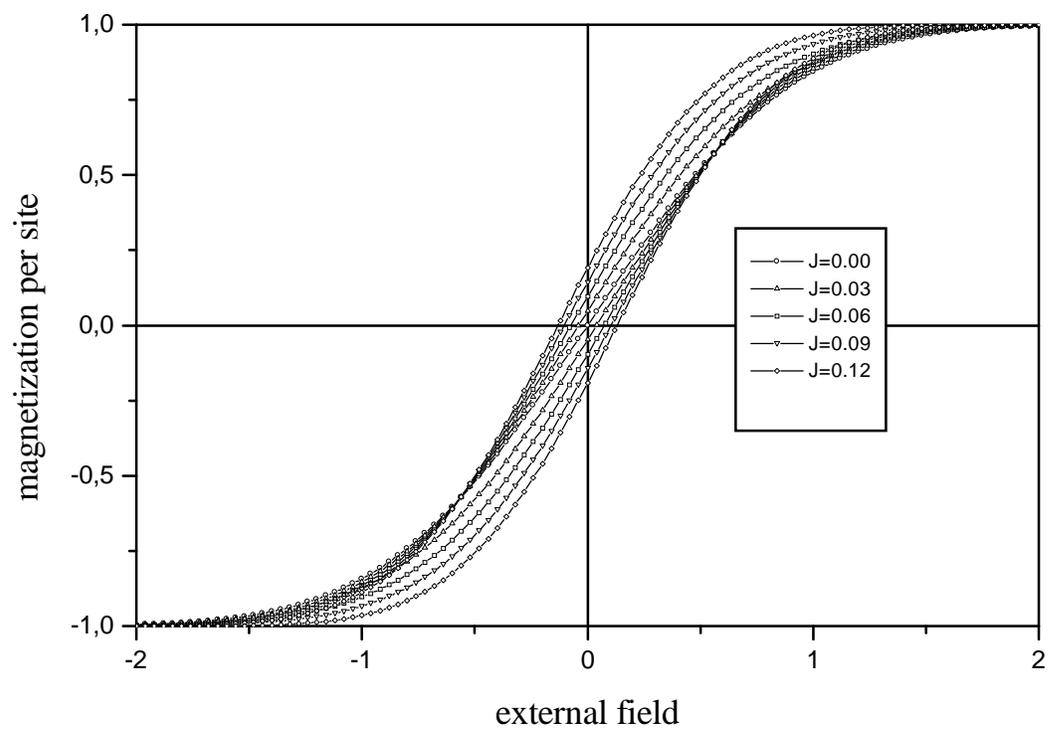
\end{document}